\documentclass[aip,reprint,english,a4paper] {revtex4-1}
\usepackage{upgreek}
\usepackage{amsfonts}
\usepackage{amstext}
\usepackage{mathtools}
\usepackage{amssymb}
\usepackage{color}
\usepackage{amsmath} 
\usepackage{graphicx} 
\graphicspath{{Figures/}{SI/}}

\newcommand{\dir}{./Bib}
\newcommand{\SMmeanradius}{S1}
\newcommand{\SMruntime}{S2}
\newcommand{\SMRotfit}{S3}
\newcommand{\SMxyz}{S4}
\newcommand{\SMCCA}{S5}
\newcommand{\SMRMSF}{S6}
\newcommand{\SMTabMosaic}{S1}

\begin{document}

\author{Daniel Nagel}
\author{Georg Diez}
\author{Gerhard Stock}
\email{stock@physik.uni-freiburg.de}
\affiliation{Biomolecular Dynamics, Institute of Physics,
	University of Freiburg, 79104 Freiburg, Germany.}
\date{\today}

\title{Accurate estimation of the normalized mutual information \\ of
  multidimensional data}

%\begin{tocentry}
%\includegraphics{\dirfig/toc.pdf}
%\end{tocentry}
\begin{abstract}

  While the linear Pearson correlation coefficient represents a
  well-established normalized measure to quantify the interrelation of
  two stochastic variables $X$ and $Y$, it fails for multidimensional
  variables such as Cartesian coordinates. Avoiding any assumption
  about the underlying data, the mutual information $I(X,Y)$ does
  account for multidimensional correlations. However, unlike the normalized Pearson correlation, it has no upper bound
  ($I \in [0, \infty)$), i.e., it is not clear if say, $I\!=\!0.4$ corresponds to a low or a high correlation. Moreover, the mutual information (MI) involves 
  the estimation of high-dimensional probability densities 
  (e.g., six-dimensional for Cartesian coordinates), 
  which requires a $k$-nearest neighbor
  algorithm, such as the estimator by Kraskov et al.\ [Phys.\ Rev.\ E
  \textbf{69}, 066138 (2004)].  As existing methods to normalize the
  MI cannot be used in connection with this
  estimator, a new approach is presented, which uses an
  entropy estimation method that is invariant under variable
  transformations. The algorithm is numerically efficient and does not
  require more effort than the calculation of the (un-normalized)
  MI. After validating the method by applying it to various toy
  models, the normalized MI between the $\text{C}_\alpha$-coordinates
  of T4 lysozyme is considered and compared to a correlation analysis
  of inter-residue contacts.

\end{abstract}
\maketitle

%
%%%%%%%%%%%%%%%%%%%%%%%%%%%%%%%%%%%%%%%%%%%%%%%%%%%%%%%%%%%%%%%%%%%%%
%

\section{Introduction}
\vspace{-4mm}

Quantifying the correlation between different observables represents
an essential aspect of data analysis across various
disciplines. Examples include geostatistics, where the
interrelationships of geographic variables in distinct spatial regions
are studied, \cite{Cliff81, Wartenberg85, Legendre94} brain research
that attempts to identify associations between various brain regions
to gain insight into the functional pathways underlying cognitive
processes,\cite{Averbeck06} computer vision, which, e.g., aligns
different images for tasks such as medical image
analysis,\cite{Viola97, Pluim03} and financial markets research that
aims to gain insight into connections, market structures, and
potential contagion effects during varying market
conditions.\cite{Barndorff04, Guo18} In the field of chemical and
biological physics, in particular, it is commonplace to study the
dynamical correlation between atoms or various parts of a molecular
system. For instance, correlation analysis is employed in principal
component analysis that reduces the dimensionality of the
system,\cite{Amadei93,Jolliffe02} in community detection approaches
that identify interacting regions of a molecule,\cite{McClendon14,
  Ravindra20, Diez22} as well as in the construction of allosteric
networks that aim to model signal transduction in
proteins. \cite{Sethi09, McClendon09,Bhattacharyya16, Wodak19}

Most commonly, correlation is measured via the Pearson correlation
coefficient $\rho$. It is defined as linear relationship between two
(in general multidimensional) variables $X$ and $Y$,
\begin{align}\label{eq:corr_1D} 
    \rho (X,Y) = \frac{\langle (X - \langle X \rangle) (Y - \langle Y
  \rangle) \rangle}{\sigma_X \sigma_Y}\;, 
\end{align}
where $\langle \ldots \rangle$ represents the statistical average and
$\sigma$ denotes the standard deviation. While the Pearson correlation
coefficient is ubiquitous in statistical analysis, it considers only
the first two moments of the underlying distribution, which---strictly
speaking---is adequate only for normally distributed data. Moreover,
the linear definition turns out to be ill-defined to account for the
correlation of multidimensional variables such as three-dimensional
(3D) Cartesian coordinates. \cite{Ichiye91,Lange06}

The concept of mutual information (MI) goes beyond these limitations,
because it avoids any assumption about the underlying
data.\cite{Cover06} Taking into account two discrete random variables
$X$ and $Y$ with their corresponding realizations
$\mathcal{X}=\{x_i\}$ and $\mathcal{Y}=\{y_i\}$, the MI measures the
statistical independence of $X$ and $Y$ by quantifying the
dissimilarity of their joint probability distribution $p_{x,y}$ and
the product of their marginal distributions $p_x p_y$ via the
Kullback-Leibler divergence of these quantities.\cite{McClendon12} This leads to the
definition
\begin{equation}\label{eq:MI_KL}
    I(X,Y) = \sum_{x\in\mathcal{X}} \sum_{y\in\mathcal{Y}}
        p_{x,y} \ln \frac{p_{x,y}}{p_xp_y}\;, 
\end{equation}
which vanishes for independent variables satisfying $p_{x,y}=p_x\, p_y$.
Introducing the Shannon entropy $H(X)$ and the joint entropy $H(X,Y)$
given by 
\begin{align}
H(X) &= -\sum_{x \in \mathcal{X}} p_x \ln p_x\;, \label{eq:entropies_discreteI}\\
H(X,Y) &= -\sum_{x \in \mathcal{X}}\sum_{y \in \mathcal{Y}}p_{x,y}
         \ln p_{x,y}\, ,    \label{eq:entropies_discreteII}
\end{align}
we may alternatively define the MI as
\begin{align}
    I(X,Y) &= H(X) + H(Y) - H(X,Y)\;, \label{eq:MI_definition} \\
  &= H(X) - H(X|Y).\label{eq:MI_intuitive}
\end{align}
Interpreting $H(X)$ as the uncertainty about $X$, and the conditional
entropy $H(X|Y) = H(X,Y) - H(Y)$ as the uncertainty about $X$
remaining after knowing $Y$, the MI describes how much the uncertainty
of $X$ decreases when we gain insight into the variable $Y$.
This intuitively appealing interpretation is readily generalized to
the treatment of multivariate data, thus rendering MI a 
robust method to identify any associations between general variables.

Nevertheless, the practical application of the MI may face two (in part
related) problems. Firstly, the computation of the MI from finite data
can be challenging, because it relies on the accurate estimation of
probability distributions $p_x$, $p_y$ and $p_{x,y}$. Given 1D data,
standard histogram or kernel-density estimation followed by numerical
integration suffices to obtain accurate estimates of the
MI. \cite{Silverman86, Scott15} The estimation of
multidimensional probability density functions, however, is notoriously
difficult and plagued by the curse of dimensionality.
For example, when we use Cartesian coordinates, the joint probability
density function is six-dimensional. As a well-established solution,
we may employ the algorithm by Kraskov, Stögbauer, and
Grassberger,\cite{Kraskov04} also known as KSG-estimator, which uses
Eq.~\eqref{eq:MI_definition} and estimates the entropies via a
$k$-nearest neighbor ($k$-nn) algorithm.\cite{Kozachenko87,
  Tsybakov96, Singh03, Lombardi16}

Secondly, many applications require a normalization of the correlation
measure. While this is naturally provided by the absolute Pearson
correlation coefficient ($|\rho| \in [0, 1]$), the MI has no upper
bound, $I(X,Y) \in [0, \infty)$. Given a specific value (say,
$I(X,Y)=0.4$), it is therefore {\em a priori} not clear whether this
corresponds to a low or a high correlation. Recalling that the MI
measures the decrease of the uncertainty of variable $X$ when we get
to know about variable $Y$ [Eq.~\eqref{eq:MI_intuitive}], it is clear
that the answer also depends on the uncertainties of the two
individual variables, given by $H(X)$ and $H(Y)$.
Hence, we need to normalize the MI to facilitate the
comparison of the MI of different sets of variables or different systems.
%or to define a dissimilarity metric via a distance $d = 1 - I \in [0, 1]$.
\cite{Studholme99, Strehl02, Ravindra20}
To this end, various strategies have been
explored. For instance, a normalized MI (NMI) can be defined by
assuming a normal joint distribution.\cite{Lange06, Gelfand59}
Alternatively, a NMI can be achieved by discretizing the data onto a
grid,\cite{Reshef11} although this might affect the statistical
robustness of the resulting measure and can be significantly slower to
compute compared to measures based on local $k$-nn
statistics.\cite{Kinney14}

In this work, we present an alternative approach to normalize the MI,
which exploits an entropy estimation method that is invariant under
variable transformations and provides an upper bound for the MI.  To
this end, we first discuss the limitations of existing methods for
estimating the NMI based on the $k$-nn approach and then describe the
key concepts underlying our new method. Using the
KSG-estimator,\cite{Kraskov04} the algorithm is highly scalable and
may handle systems of high dimension. We validate our approach by
applying it to various toy models, and compare to previous
estimators.
As a real-world example, we consider molecular dynamics simulation
data of T4 lysozyme (T4L), a bistable protein consisting of $162$
residues.\cite{Dixon92,Ernst17,Post22a} To explain the functional
motion of T4L, we calculate the NMI between the Cartesian coordinates
of all $\text{C}_\alpha$-atoms, and compare the results to a MoSAIC
\cite{Diez22} correlation analysis of the inter-residue contacts of
T4L. Our results demonstrate the effectiveness of our approach in
accurately estimating the NMI of multidimensional variables.

%\newpage
%
%%%%%%%%%%%%%%%%%%%%%%%%%%%%%%%%%%%%%%%%%%%%%%%%%%%%%%%%%%%%%%%%%%%%%
%
\section{Theory and methods}
\subsection{Normalizing the mutual information}
\label{sec:NormalizingMI}
\vspace{-4mm}

To motivate a suitable normalization factor, we consider
the well-known inequalities\cite{Vinh10}
\begin{align}
  I(X,Y) &\le \min\limits_{Z=X, Y} H(Z) \le \sqrt{H(X) H(Y)}
           \nonumber \\
  &\le \max\limits_{Z=X, Y} H(Z) \le H(X,Y).
    \label{eq:upperlimits}
\end{align}
Aiming at a normalization that depends on both random variables, and
since $I(X,X) = H(X)$, we choose the geometric mean because of its
analogy to a normalized inner product and thus to the Pearson
correlation coefficient.\cite{Strehl02} This leads to the upper bound
of the NMI defined as
\begin{align} \label{eq:MI_normalization}
    I_\text{N}(X,Y) = \frac{I(X,Y)}{\sqrt{H(X)H(Y)}} \leq 1.
\end{align}

Unfortunately, however, the above inequalities hold only for discrete
entropies but not for continuous entropies.\cite{Jaynes68} This poses
a problem, as the KSG-estimator of the NMI is based on continuous
$k$-nn statistics, which can violate Eq.~\eqref{eq:upperlimits}. To
facilitate the use of the inequalities, the discrete probability
distributions (such as $p_x$) in the entropy definitions in
Eqs.~\eqref{eq:entropies_discreteI} and
\eqref{eq:entropies_discreteII} must therefore be converted to
continuous distributions (such as $P(x)$). With this end in mind,
Shannon\cite{Shannon48} suggested to use the continuous form
\begin{align}\label{eq:entropy}
    H_{\rm d}(X) = - \int_\mathcal{X} \mathrm{d} x P(x) \ln P(x),
\end{align}
referred to as 'differential' entropy. From the normalization condition
\begin{align}\label{eq:entropy2}
1 = \int_\mathcal{X} \mathrm{d}x\, P(x),
\end{align}
it is obvious that the density $P(x)$ carries the dimension $1/[x]$, which
causes a problem when calculating the entropy via
$\ln{P(x)}$. Moreover, the normalization condition Eq.~\eqref{eq:entropy2}
may result in negative entropy estimates. To see this, we first
consider the discrete case with normalization
$1 \overset{!}{=} \sum_{x \in \mathcal{X}}p_x$, where every event
occurs with a probability between zero and one (i.e., $p_x \le 1$),
which results in $\ln p_x \le 0$ and ensures that the entropy is
positive valued. In the continuous case, on the other hand, the
probability density is normalized in Eq.~\eqref{eq:entropy2} such
that the area under its curve is one, which has the consequence that
$P(x)$ can take any value in $[0, \infty)$. Thus,
$\ln P(x) \in (-\infty, \infty)$, which results in an entropy that is
not necessarily positive.

Jaynes\cite{Jaynes68} showed that the root cause of this problem lies
in the lack of invariance of Eq.~\eqref{eq:entropy} under a variable
transformation $x \rightarrow \Tilde{x}$. He derived the correct
continuous limit of the Shannon entropy, \cite{note0nmi} referred to as
'relative' entropy
\begin{align} \label{eq:relative_entropy}
    H_{\rm r}(X) = - \int_\mathcal{X} \mathrm{d}x\, P(x) \ln \frac{P(x)}{m(x)},
\end{align}
where he introduced the invariant measure $m(x)$ that transforms
identically to $P(x)$, i.e., 
\begin{equation}
    x\to \tilde{x}:\quad\frac{P(x)}{m(x)} = \frac{P(\tilde{x})}{m(\tilde{x})}.
\end{equation}
By design, the relative entropy $H_{\rm r}(X)$ does not depend on the
choice of coordinates or units.

Note that the invariant measure of two coordinates can be chosen to
factorize in single-coordinate functions, i.e., 
\begin{align} \label{eq:m_factorisation}
    m(x,y) = m(x)m(y) ,
\end{align}
because $m(x,y)$ must transform both variables like the respective
coordinate. As a consequence, the MI remains invariant upon the
introduction of the invariant measure, 
\begin{align}
    I_{\rm d}(X,Y) &= H_{\rm d}(X) + H_{\rm d}(Y) - H_{\rm d}(X,Y)\nonumber\\
                   &= H_{\rm r}(X) + H_{\rm r}(Y) - H_{\rm r}(X,Y)
+ \int \mathrm{d}(x, y)
                     \nonumber\\
    &\times P(x,y)
        \underbrace{\left[\ln m(x) + \ln m(y) - \ln m(x)m(y)\right]}_{=0}\nonumber\\
    &= I_{\rm r}(X,Y) ,
    \label{eq:MI_invariance}
\end{align}
rendering the differential (subscript 'd') and relative (subscript
  'r') definitions equivalent. It is important to stress, however,
that this equivalence does not hold for the entropies, i.e.,
$H_{\rm d}(X) \ne H_{\rm r}(X)$, and therefore also not for the NMI in
Eq.\ \eqref{eq:MI_normalization}. Hence, it is crucial to apply
relative entropies [Eq.~\eqref{eq:relative_entropy}] when we compute
the normalization between two continuous random variables. Otherwise,
the usage of different representations or parametrizations of the two
probability densities could lead to different results for the NMI,
which renders it challenging to compare or interpret the results.

We note in passing that various alternative definition of the NMI
exist. Obviously, we may simply divide the MI by its maximum $I_\text{max}$, 
i.e., 
\begin{align} \label{eq:MI_normalizationB}
    I_\text{NM}(X,Y) = \frac{I(X,Y)}{I_{\rm max}}.
\end{align}
Moreover, we can use the joint entropy from
Eq.~\eqref{eq:upperlimits} and define
\begin{align} \label{eq:MI_normalizationC}
    I_\text{NJ}(X,Y) = \frac{I(X,Y)}{H(X,Y)}.
\end{align}
Another approach is to employ the transformation of Gel’fand and
Yaglom, \cite{Gelfand59,Lange06} i.e.,
\begin{align}
I_\text{GY}(X,Y) = \sqrt{1 - \exp[-2 I(X,Y) / (d_X + d_Y)]}, \label{eq:Igy}
\end{align}
where $d_X$ and $d_Y$ denote the dimension of variables $X$ and $Y$,
respectively. The definition of $I_\text{GY}(X,Y)$ clearly provides a
mapping from the $I(X,Y) \in [0, \infty)$ to a normalized
quantity.
While this can be achieved by various mappings, $I_\text{GY}$ can
  be shown to be identical to the linear correlation $|\rho|$ in the
  special case that $P(x,y)$ is a bivariate normal distribution. In
the general case, however, the interpretation of $I_\text{GY}$ as a
correlation measure is less clear, because it does not take into
  account the upper bound of the mutual information based on the
  individual entropies of $H(X)$ and $H(Y)$, see Eq.\
  (\ref{eq:upperlimits}). In particular, $I_\text{GY}$ shows
relatively large values already for weakly correlated
data.\cite{Diez22} 
%
%%%%%%%%%%%%%%%%%%%%%%%%%%%%%%%%%%%%%%%%%%%%%%%%%%%%%%%%%%%%%%%%%%%%%
%
\subsection{KSG-estimator using the continuous entropy}
\vspace{-4mm}

While various continuous entropy estimators have been
proposed,\cite{Nemenman01, Jiao15, Archer14, Valiant17} we wish to
adapt the KSG-estimator\cite{Kraskov04} to compute the NMI, as it is
able to treat multidimensional problems. The algorithm is based on
local $k$-nn statistics, that is, it aims to calculate the probability
distribution $p_i$ of the distance between the data points $\{x_i\}$
and their $k$th nearest neighbor, assuming a local constant density.
This probability can be visualized as the volume of a $d$-dimensional
sphere with radius $2\varepsilon_i$, which is a proxy for the density
of the points around $x_i$.
The entropy is then given as the
expectation value of $-\ln p_i$, which can be calculated from the
digamma function $\psi$. For further reference, we only mention here
the main results of the algorithm, and refer to the original paper for
details.\cite{Kraskov04}

Kraskov et al.\cite{Kraskov04} used the multinomial theorem to calculate the
probability $p_i$ and averaged the log-likelihood over all particles
to obtain an estimate of the entropy.
Starting with the definition of differential entropy in Eq.\
\eqref{eq:entropy}, they derived unbiased
estimators $\hat{H}_{\rm d}$ for the differential entropy and the
joint entropy
\begin{align}
  \hat{H}_{\rm d}(X) = &-\langle\psi(n_x +1)\rangle + \psi(N)  +
                       \ln c_{d_X} \nonumber \\
                       &+ d_X \langle\ln 2 \varepsilon \rangle,
    \label{eq:KSG_entropy_estimator} \\
%    \intertext{and for the joint entropy}
    \hat{H}_{\rm d}(X,Y) = &- \psi(k) + \psi (N) + \ln
                           (c_{d_X}c_{d_Y}) \nonumber \\
                           &+ (d_X + d_Y)\langle \ln 2 \varepsilon\rangle .
    \label{eq:KSG_joint_entropy_estimator}
\end{align}
Here the first two terms of $\hat{H}_{\rm d}$ represent the average
log-likelihood of finding the $k$th nearest neighbor distance
$\varepsilon$,
which can be expressed by the digamma function $\psi(n)$ that
satisfies the recursion $\psi(n+1) = \psi(n) + 1 /n$, with the
Euler-Mascheroni constant as starting value
$\psi(0)=-\gamma \approx 0.577$. Furthermore, $N$ denotes the number
of data points, $\langle \ldots \rangle$ represents the average over
these points, $k$ is the number of nearest neighbors considered,
$\varepsilon$ is the distance between a considered data point and its
$k$th nearest neighbor.  Moreover, $n_x$ is the number of points whose
distance in $x$-direction from the considered data point is less than
$\epsilon$ (analogously for $n_y$) and $c_{d}$ is the volume of the
$d$-dimensional unit ball. \cite{note4nmi}
Hence the latter two terms of $\hat{H}_{\rm d}$ account for the average
  log-volume of the $k$th nearest neighbor sphere.
Inserting Eqs.~\eqref{eq:KSG_entropy_estimator} and
\eqref{eq:KSG_joint_entropy_estimator} into
Eq.~\eqref{eq:MI_definition}, we obtain the KSG-estimator of the MI
\begin{align} \label{eq:KSG_differential}
    I(X,Y) = \psi(N) + \psi(k) - \langle \psi(n_x \!+\! 1) +
  \psi(n_y \!+\!1 ) \rangle. 
\end{align}

Using the differential entropy, the above estimator works perfectly
fine to compute the (non-normalized) MI, since it is invariant with
respect to a variable transformation due to
Eq.~\eqref{eq:MI_invariance}. To calculate the normalized MI in
Eq.~\eqref{eq:MI_normalization}, however, we need to ensure that the
entropies $H(X)$ and $H(Y)$ are also invariant. Instead of the
differential entropy, we therefore use the relative
entropy estimator $\hat{H}_{\rm r}$, which effectively introduces an
additional term containing the invariant measure $m(X)$,
\begin{align}
    \hat{H}_r(X) = - \left\langle\ln \frac{p(X)}{m(X)}\right\rangle
  = \hat{H}_{\rm d}(X) + \langle \ln m(X)\rangle
%  \nonumber\\
%    &=-\langle\psi(n_x+1)\rangle + \psi(N) + \ln c_d + d_x \langle \ln
%      2\varepsilon_i\rangle + \langle \ln m(X)\rangle \;,
\end{align}
and analogously for $\hat{H}_r(X,Y)$. This results in the estimators
\begin{align}
    \hat{H}_r(X)
        = &- \langle \psi(n_{x} + 1)\rangle + \psi(N) + \ln c_{d_X} \nonumber\\
        &+ d_X
        \langle \ln 2 \varepsilon\rangle + \langle\ln m(X)\rangle,
          \label{eq:Hc_x_estimator}\\
  \hat{H}_r(X,Y)
        = &- \psi(k) + \psi (N) + \ln (c_{d_X} c_{d_Y}) \nonumber\\
        &+ (d_X +
          d_Y)\langle \ln 2 \varepsilon\rangle + \langle\ln m(X,Y)\rangle,
        \label{eq:Hc_xy_estimator}
\end{align}
which now additionally depend on the invariant measures $m(X)$ and 
$m(X,Y)$.

%
%%%%%%%%%%%%%%%%%%%%%%%%%%%%%%%%%%%%%%%%%%%%%%%%%%%%%%%%%%%%%%%%%%%%%
%
\subsection{Estimating the Invariant Measures}
\vspace{-4mm}

Having ensured scale and parametrization invariance of the
KSG-estimator via the introduction of invariant measures, the next
step is to calculate these measures. To avoid bias,
Jaynes\cite{Jaynes68} suggested using an invariant measure $m(X)$
that assumes complete ignorance, i.e., a constant probability
density. For 1D data $\{ x_i \}$ distributed between $a$ and $b$, this
yields
\begin{align} \label{eq:jaynes_assumption}
    1 = \int_a^b m \;\mathrm{d} x\,  \Rightarrow m
    = \frac{1}{a-b} , %= \text{const.}
\end{align}
that is, the invariant measure is constant and defined via the
boundaries of the data set. Although this ansatz makes sense for 1D
problems, in higher dimensions it amounts to a uniform sampling of the
data within a $d$-dimensional cube, which is hardly appropriate for
finite data. In particular, data outliers may lead to a massive
overestimation of the volume and thus to a significant underestimation
of the density, resulting in an invalid estimate of the entropy.

To address this issue, we propose a new estimator of the invariant measure.
%Similar as in Eq.~\eqref{eq:jaynes_assumption},
Following Jaynes' suggestion,\cite{Jaynes68} we assume that the 
measures $m(X)$ and $m(X,Y)$
%with $X \in \mathbb{R}^{d_X}$ and $Y \in \mathbb{R}^{d_Y}$ $d_X+d_Y$-dimensional
are the inverse of the corresponding volumes enclosed by the data points, i.e., 
\begin{align}
    m(X) &= 1/V(X) ,\\ %= m(\boldsymbol{\varepsilon}, \boldsymbol{n}_x)
    m(X,Y) &= 1/V(X,Y).% m(\boldsymbol{\varepsilon}, \boldsymbol{n}_x, \boldsymbol{n}_y)
\end{align}
To achieve an efficient estimation of the volumes, we only use
quantities that are already computed in the original KSG-algorithm. As
introduced in
Eqs.~\eqref{eq:KSG_entropy_estimator}--\eqref{eq:KSG_differential},
these include the distances $\varepsilon$ between the data points and
their $k$-th nearest neighbors, the number of points $n$ whose
distance from a considered data point is less than $\epsilon$, and the
volume $c_{d}$ of the $d$-dimensional unit ball. To estimate the
volume $V(X,Y)$, we calculate $N$ times the mean volume of a single
data point $\langle (2 \varepsilon)^{d_{X} + d_{Y}} \rangle$, and
divide it by the number of nearest neighbors $k$, to avoid
overcounting. This yields the estimator
\begin{align}\label{eq:volume_v}
    \hat{V}(X,Y) &= \frac{N}{k} c_{d_{X}} c_{d_{Y}} \langle (2
           \varepsilon)^{d_{X} + d_{Y}} \rangle.
\end{align}
To ensure the factorization of the measure
[Eq.~\eqref{eq:m_factorisation}], we request that
\begin{align}
    \hat{V}(X,Y) = \hat{V}(X) \hat{V}(Y) ,
\end{align}
which leads directly to the invariant measures \cite{note5nmi}
\begin{align}
    \ln \hat{m}(X) &= - \ln c_{d_X} - \frac{d_X}{d_X + d_Y} \ln \langle (2\varepsilon)^{d_X + d_Y}
               \rangle ,\label{eq:m_X} \\
    \ln \hat{m}(X, Y) &= - \ln (c_{d_X} c_{d_Y}) - \ln \langle
                  (2\varepsilon)^{d_X + d_Y} \rangle.\label{eq:m_v} 
\end{align}
Inserting these expressions into
Eqs.~\eqref{eq:Hc_x_estimator} and \eqref{eq:Hc_xy_estimator}, we
obtain as a final result the entropy estimators
\begin{align}
    \hat{H}_r(X) %&= H(X) + \ln m(X)\nonumber\\
    &= - \langle \psi(n_{x} + 1)\rangle + \psi(N) + d_X \langle \ln
      \tilde{\varepsilon}\rangle, 
    \label{eq:Hcx_estimator}\\
    \hat{H}_r(X, Y) %&= H(X,Y) + \ln m(X,Y)\nonumber\\
    &= - \psi(k) + \psi(N) + (d_X + d_Y) \langle \ln
      \tilde{\varepsilon}\rangle, 
    \label{eq:Hcxy_estimator}
\end{align}
where $\tilde{\varepsilon} = \varepsilon /\langle
\varepsilon^{d_X + d_Y}\rangle^{1/(d_X + d_Y)}$
%\begin{align}
%    \tilde{\varepsilon_\alpha} &=\frac{\varepsilon}{\sqrt[d_\alpha]{\langle
%                                 \varepsilon^{d_\alpha}\rangle}} \;, 
%\end{align}
denotes the scaling-invariant $k$-nn radius. Equations
\eqref{eq:m_X}--\eqref{eq:Hcxy_estimator} represent the main
theoretical results of this paper. As the inequalities in
Eq.~\eqref{eq:upperlimits} apply for the above entropy estimators
$\hat{H}_r$, the NMI can be obtained through the normalization factor
$(\hat{H}_r(X)\hat{H}_r(Y))^{-\frac{1}{2}}$ as in
Eq.~\eqref{eq:MI_normalization}, or alternatively via the factor
$\hat{H}_r(X,Y)^{-1}$ as in Eq.~\eqref{eq:MI_normalizationC}.
We note that Eqs.\ \eqref{eq:m_X}--\eqref{eq:Hcxy_estimator} address the
  two main challenges in computing MI: accurately estimating
  probability distributions in high-dimensional spaces with finite
  data, and normalizing MI for meaningful comparisons across different
  data sets. The first challenge is addressed by using $k$-nearest
  neighbor statistics to approximate probability distributions; the
  second one by normalizing MI using relative entropy, which is
  achieved by introducing an invariant measure that assumes a constant
  density.

Apart from the mean volume-based approach discussed above, we also
derived and discussed a mean radius-based algorithm for the volume
estimator. It is defined as $N$ times the volume of a
$d$-dimensional sphere $c_d$ (with the average radius of the $k$th
nearest neighbor) divided by $k$, see the SI. While both algorithms
are quite similar, the radius-based estimator suffers from a
systematic bias for small values of $k$ caused by boundary effects,
leading to an underestimation of the volume, see Fig.~\SMmeanradius{}
in the Supplementary Material.

%\newpage
%
%%%%%%%%%%%%%%%%%%%%%%%%%%%%%%%%%%%%%%%%%%%%%%%%%%%%%%%%%%%%%%%%%%%%%
%
\subsection{Validation}
\label{sec:ToyModel}
\vspace{-4mm}

\begin{figure}[ht!]
  \includegraphics{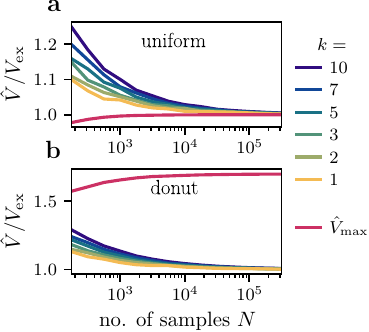}
  \caption{Estimation of the volume of (a) a
    uniform distribution ($x, y\in [0, 1]$) and (b) uniformly
    distributed data on a donut-shaped area with $r=\sqrt{x^2+y^2}\in[0.5, 1]$.
    Shown as a function of the sample size $N$, results of the $k$-nn estimator
    [Eq.~\eqref{eq:volume_v}] for various numbers $k$ of nearest
    neighbors are compared to the boundary estimate $\hat
    V_\text{max}$ [Eq.~\eqref{eq:jaynes_assumption}]. 
%    For each value of $N$, we average over 100 calculations of $\hat V$.
    \label{fig:validation1}}
\end{figure}

Let us first test the accuracy of the volume estimator in
Eq.~\eqref{eq:volume_v} for simple example distributions. To this end,
Fig.\ \ref{fig:validation1} shows estimations of $V(X)$ as a function
of the sample size $N$, obtained for various numbers of nearest
neighbors $k$. For easy representation, we divide $\hat V(X)$ by the
exact volume $V_{\rm ex}(X)$, such that the ratio approaches 1 for
large $N$.  Moreover, we average over 100 calculations of $\hat V$, in
order to suppress large fluctuations of $\hat V(X)$ for small $N$. In
the case of a 2D uniform distribution ($x, y\in [0, 1]$), we find that
the $k$-nn estimator converges rapidly to the correct result, and
depends (for not too small $N$) only little on $k$ (Fig.\
\ref{fig:validation1}a). We also estimate the volume from the
boundaries of the data set according to Eq.\
(\ref{eq:jaynes_assumption}), i.e.,
$\hat V_\text{max} = (x_{\rm max}-x_{\rm min})(y_{\rm max}-y_{\rm
  min})$. As expected, the latter ansatz works perfectly well for a
uniform distribution.
The situation is different in our second example, where the data are
uniformly distributed in a donut-shaped area (Fig.\
\ref{fig:validation1}b). While the boundary-based estimation again
gives the full 2D area (which in the present example is too large by
$\sim 70\,\%$), the $k$-nn estimator reliably converges to the
correct result.

%
%%%%%%%%%%%%%%%%%%%%%%%%%%%%%%%%%%%%%%%%%%%%%%%%%%%%%%%%%%%%%%%%%%%%%
%
%\newpage
Next we wish to assess the accuracy and performance of the new
estimator of the NMI defined in Eq.~\eqref{eq:MI_normalization}.
%using an analytically solvable 2D model that allows for reference
%calculations of the entropy. The model consists of two Gaussian
To this end, we employ a 2D toy model with the distribution
\begin{equation} \label{eq:2Dmodel}
  P(r = \sqrt{x^2+y^2}) \propto \exp \left[ - \frac{(r - r_0)^2}{2\sigma^2} \right],
\end{equation}
with mean $r_0=0.75$ and standard deviation $\sigma = 1 / 8$.
%, resulting in the 2D free energy landscape $\Delta G(x, y) = -k_{\rm B}T \ln P$.
In the following, we compare the results of our $k$-nn estimator $I_{\rm N}$
with $k=5$ to other commonly used methods.

We first compare the results for $I_{\rm N}$ to the most commonly used
method, that is, the binning of the variables. This discrete
estimation of the NMI $I_{\rm N, hist}$ appears rather simple and is
expected to yield good results for 2D data.\cite{Ravindra20} It should
be stressed, though, that for finite data, the entropy estimates (and
thus the NMI) depend considerably on the number of bins
$N_\text{bins}$ used per dimension. Comparing the NMI obtained for
various choices of $N_\text{bins}$ and sample sizes $N$,
Fig.~\ref{fig:validation2}a shows that the accuracy of the binning
approach indeed depends critically on these parameters. While most of
the choices clearly overestimate the NMI, we find that the previously
suggested heuristic of selecting the number of bins to minimize the
NMI achieves the closest approximation to our $k$-nn estimator
result.\cite{Ravindra20}

%\newpage

Generally speaking, the assumption of a uniform probability
distribution $P(x) = N_\text{bins}^{-1}$ leads to
$H(X) = \ln N_\text{bins}$, which diverges as $N_\text{bins}$
approaches infinity. For finite sampling and bin numbers, on the other
hand, the entropy is maximized once there is only a single point in
each bin. This limit leads to $P(x) = 1/N = P(x, y)$ and thus to an
NMI of 1.  Defining $H_\text{max}$ as the largest possible entropy for
a given number of samples $N$, we thus find that both estimators,
$N_\text{N}$ and $N_\text{N,hist}$, exhibit a $1 / H_\text{max}$
scaling with $N$.
When we multiply the results for the $k$-nn estimator $I_{\rm N}$ and
for the binning method $I_{\text{N,hist}}$ with $H_\text{max}$,
Fig.~\ref{fig:validation2}b indeed reveals that the resulting
quantities depend only minor on $N$. Due to the simplicity of the
model, the fluctuations of the NMIs (shaded areas) become smaller than
the line width for $N \gtrsim 10^3$.

To demonstrate the necessity of introducing an invariant measure to
calculate $I_{\rm N}$, Fig.~\ref{fig:validation2}b also shows results for
$I_{\rm N,d}$, which were obtained by directly using the differential
entropy [Eq.\ (\ref{eq:entropy})] for the NMI
estimator.\cite{Hernandez17}. Getting negative values for the NMI in
our model example, this is clearly not an option.
Fig.~\ref{fig:validation2}b also shows the NMI of Gel’fand and Yaglom
[{Eq.\ (\ref{eq:Igy})], which is significantly larger than the values
  obtained from the $k$-nn estimator and for the binning method.

\begin{figure}[ht!]
  \includegraphics{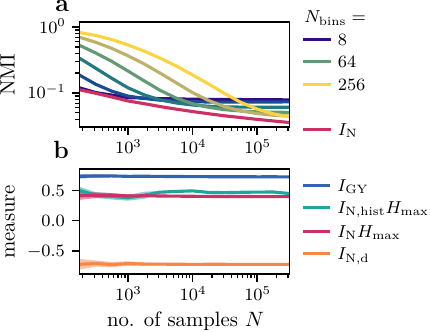}
  \caption{Estimation of the NMI for the 2D model defined in Eq.~\eqref{eq:2Dmodel}, shown as a function of the sample size $N$.
    (a) Comparison of $I_{\rm N}$ obtained from the $k$-nn estimator to the discrete estimation of the NMI $I_{\rm N, hist}$ obtained via 2D binning of the data using various numbers of bins $N_{\rm{bins}}\in[8,256]$. 
    (b) Comparison of various estimations for the NMI, indicating the respective standard deviation by a shaded area. Shown are the $k$-nn estimator $I_{\rm N} H_\text{max}$ ($k=5$), the 2D binning result $I_{\rm N, hist} H_\text{max}$  ($N_{\rm{bin}}\in[8, 256]$), the NMI $I_\text{GY}$ of Gel’fand and Yaglom [Eq.~\eqref{eq:Igy}], and $I_{\rm N,d}$ using the differential entropy [Eq.~\eqref{eq:KSG_differential}]. Here $H_\text{max}$ corresponds
    to the highest possible entropy (i.e., the entropy of a uniform
    distribution for a given number of samples $N$).
    For each value of $N$, we average over 100 calculations.
    \label{fig:validation2}}
\end{figure}

Finally, we briefly discuss the effort to compute the proposed
estimators. The estimator scales with the number of samples $N$ as
$\mathcal{O}(N\ln N)$, due to the $k$-d tree used for the nearest
neighbor search. With respect to the number of coordinates $M$, we get
the obvious scaling of $\mathcal{O}(M^2)$. We wish to stress that the
NMI comes at the same price as the MI, since the ingredients of the
normalization factor (e.g., $H(X)$) are calculated anyway.
Let us mention some runtimes examples for the new NMI
estimator, using an Intel\textsuperscript{\textregistered{}}
Core\texttrademark{} i9-10900 CPU. For $N=10^5$ samples and 230 1D
coordinates, the computation of the $230 \times 230$ NMI matrix takes
about 1 CPU hour.  A similar time is needed to calculate the NMI
matrix for $N=10^4$ samples and 190 3D coordinates. For further
details, see Fig.~\SMruntime.

%\newpage
%
%%%%%%%%%%%%%%%%%%%%%%%%%%%%%%%%%%%%%%%%%%%%%%%%%%%%%%%%%%%%%%%%%%%%%
%
\section{Application: Functional dynamics of T4L}
\vspace{-4mm}
\subsection{Model system}
\vspace{-4mm}

As an interesting and non-trivial example, we apply the new algorithm
to explain the interresidue correlations associated with the
functional dynamics of T4 lysozyme (T4L). \cite{Dixon92,
Ernst17, Post22a} Aiming to destroy bacterial cell walls by
catalyzing the cleavage of glycosidic bonds, T4L performs an
open-closed transition of its two domains that resembles a Pac-Man,
see Figs.\ \ref{fig:T4L}a,b. This motion of the ``mouth'' region was
recently shown to be triggered by a locking mechanism in the ``hinge''
region of T4L, by which the side chain of Phe4 changes from a
solvent-exposed to a hydrophobically-buried state. \cite{Ernst17} To
elucidate the mechanism underlying this long-range (or allosteric)
coupling between the two distant regions, various dimensionality
reduction approaches have been employed.\cite{Hub09,Lange08a,
  Ernst17,Brandt18,Sittel18} Applying the MoSAIC\cite{Diez22}
correlation analysis to all interresidue contacts of T4L, a network of
interresidue distances was identified (Fig.\ \ref{fig:T4L2}a), which
move in a concerted manner.\cite{note3nmi} The cooperative process
originates from a cogwheel-like motion of the hydrophobic core in the
hinge region, which constitutes a flexible transmission
network. Through rigid contacts and the protein backbone, the small
local changes of the hydrophobic core are passed on to the distant
terminal domains and lead to the emergence of a rare global
conformational transition.

Given this exceptionally well-studied example of allosteric coupling,
it is instructive to further characterize the underlying
conformational transition by considering the NMI of the
C$_\alpha$-atom coordinates. The resulting correlation matrix
constitutes a residue interaction network, which is commonly employed
to reveal allosteric pathways using network theory
methods. \cite{Sethi09, McClendon09, Bhattacharyya16, Wodak19} To this
end, we adopt the $50\,\mu$s-long all-atom molecular dynamics (MD)
simulation by Ernst et al.\cite{Ernst17}, which employed Gromacs
4.6.7\cite{Hess08} in combination with the Amber ff99SB*-ILDN force
field\cite{Hornak06, Best09, LindorffLarsen10} and the TIP3P water
model.\cite{Jorgensen85}

%Using $N=5\cdot 10^5$ data points and $k=5$ next neighbors for the
%KSG-estimator, the calculation of the NMI between the 164 residues of
%T4L takes approximately $40\,$hours on an
%Intel\textsuperscript{\textregistered{}} Core\texttrademark{}
%i9-10900 CPU.

Using Cartesian C$_\alpha$-atom coordinates, first the translation and
overall rotation must be removed from the MD trajectory. Since the
rotation depends via the moment of inertia on the molecule’s
structure, this separation is only straightforward for
rigid systems, but can lead to serious artifacts when we calculate the
linear correlation matrix [Eq.\ (\ref{eq:corr_1D})] of molecules that
undergo large-amplitude motion (as, e.g., folding
proteins). \cite{Sittel14} Although being scale-invariant with respect
to linear coordinate transformations,\cite{Cover06} this problem also
occurs for the MI. To test if this is an issue when considering the
open-closed motion of T4L, we performed a
global rotational fit (where both open and closed conformations are
rotated with respect to a common reference structure) and a local
rotational fit (where the open and closed conformations are rotated
with respect to their respective reference structure). Comparing the
resulting estimates of the linear correlation and the NMI, Fig.\ \SMRotfit{}
reveals only minor differences, suggesting that the standard
rotational fit procedure works satisfactorily to describe the functional
dynamics T4L.

%\newpage
%
%%%%%%%%%%%%%%%%%%%%%%%%%%%%%%%%%%%%%%%%%%%%%%%%%%%%%%%%%%%%%%%%%%%%%
%
\subsection{Comparison of correlation measures}
\label{sec:T4LInt}
\vspace{-4mm}

The $50\,\mu$s-long MD trajectory of T4L populates the open and the
closed state for $\sim70$\% and 30\,\% of the time, respectively, and
exhibits mean waiting times of
$\tau_{\rm o\rightarrow c} \sim 4\,\mu$s and
$\tau_{\rm c\rightarrow o} \sim 2\,\mu$s, which are in excellent
agreement with recent experimental results.\cite{Sanabria20} 
%
%Furthermore, we excluded all transitions between the two metastable
%conformations to mimic equilibrium simulations in which sampling
%the transition is computationally out of reach.
%
Applying the new algorithm to the MD data,
Fig.~\ref{fig:T4L}c shows the
resulting NMI matrix $I_{\rm N}$ obtained for the C$_\alpha$-atom
coordinates. As may be expected, we find large values of $I_{\rm N}$
along the diagonal, revealing strong correlations with the first
next-neighbor residues. In well-defined secondary structures such as
the $\alpha$-helices this extends typically to 3--4 next neighbors, in
the connecting loops only to 1--2.
Moreover, the matrix roughly splits up in four blocks, which are
defined by the N-domain ($\alpha_1$--$\alpha_3$, $\beta_1$--$\beta_3$)
and the C-domain ($\alpha_5$--$\alpha_9$), see Fig.\
\ref{fig:T4L}a. While the diagonal blocks account for the internally
quite rigid two domains, the less prominent off-diagonal blocks
indicate correlated motions of the N- and C-domain. The
$\alpha_4$-helix between the two domains couples only weakly to the
adjacent helices and therefore acts as a lowly correlated buffer
zone. This is because it lacks polar contacts with neighboring
helices. %, which adds to the flexibility of $\alpha_4$.
At the C-terminal end, the $\alpha_{10}$-helix is considerably exposed
to the solvent and therefore correlates less with the core of the
protein.

\begin{figure}[t!]
    \centering
    \includegraphics[width=0.9\linewidth]{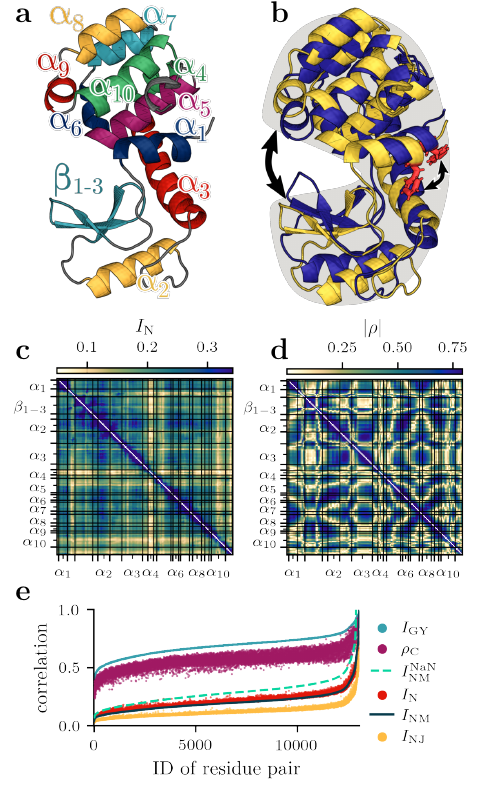}
    \caption{ Correlation analysis of the functional dynamics of T4
      Lysozyme, using the Cartesian coordinates of all $C_\alpha$
      atoms.  Molecular structure of T4L, with panel (a) indicating
      the secondary structures, and panel (b) showing the open
      (yellow) and closed (blue) conformations. The open-closed motion
      of the mouth region by a black arrow and the locking mechanism
      in the hinge region by residue Phe4 (red).  (c) $k$-nn
      estimation [Eq.~\eqref{eq:MI_normalization}] of the NMI and (d)
      corresponding absolute Pearson correlation coefficient [Eq.\
      (\ref{eq:corr_1D})]. (e) Comparison of various correlation
      measures, displayed in the order of increasing values of the
      (un-normalized) MI of all residue pairs. Shown are our standard
      definition $I_{\rm N} = I/\sqrt{H(X)H(Y)}$, the variant using
      the joint entropy $I_\text{NJ}(X,Y) = I(X,Y)/H(X,Y)$, the
      maximum-normalized $I_\text{NM}(X,Y) = I(X,Y)/I_{\rm max}$, the
      maximum-normalized $I_\text{max}^\text{NaN}$ that neglects
      next-neighbor correlations,
      $I_\textsc{GY} = \sqrt{1-\exp (-I/3)}$ according to Gel'fand and
      Yaglom, and the canonical correlation $\rho_\text{C}$.}
    \label{fig:T4L}
\end{figure}

It is interesting to compare the above results for the NMI to the linear 
Pearson correlation coefficient defined in Eq.\ (\ref{eq:corr_1D}). 
Choosing the variables $X=(x_i,y_i,z_i)^{\rm T}$ and
$Y=(x_j,y_j,z_j)^{\rm T}$ for the C$_\alpha$-coordinates of residues
$i$ and $j$, respectively, the absolute Pearson coefficient is given
as the sum of three directional components with $\alpha=x,y,z$
\begin{equation} \label{eq:Person3D}
|\rho_{ij}| = \Big| \sum_\alpha \rho_{ij}^{\alpha \alpha} \Big| .
\end{equation}
That is, due to the scalar product in Eq.\ (\ref{eq:corr_1D}), only
diagonal elements of $\rho_{ij}^{\alpha \beta}$ with
$\alpha \!=\! \beta$ occur. Comparing the results for the Pearson
coefficient and the NMI, Figs.\ \ref{fig:T4L}c and \ref{fig:T4L}d show
that the two correlation measures differ substantially, i.e., the
structure of the Pearson matrix shows prominent patterns that are
absent in the NMI. This artifact of the Pearson coefficient is due to
the fact that the components $\rho_{ij}^{\alpha \alpha}$ at the same
time account for the correlation between residues $i$ and $j$ and the
relative orientation of the directions $\alpha=x,y,z$. Since the
components can take positive and negative values
($-1 \le \rho_{ij}^{\alpha \alpha} \le 1$), for example, a large
positive value for $\rho_{ij}^{xx}$ can be canceled by a large
negative value for $\rho_{ij}^{yy}$, although residues $i$ and $j$ are
highly correlated. Showing the matrices of the three components
$\rho_{ij}^{xx}$, $\rho_{ij}^{yy}$, and $\rho_{ij}^{zz}$, Fig.\
\SMxyz{} readily explains the spurious pattern of the Pearson
correlation matrix. Although this failure of the Pearson measure for
Cartesian coordinates was previously pointed out by Lange and
Grubm\"uller, \cite{Lange06} the Pearson correlation coefficient
nevertheless has seen widespread use in the construction of allosteric
networks.\cite{Sethi09, McClendon09, Bhattacharyya16, Wodak19}

As a simple remedy, we may sum the moduli of the three components,
i.e., $|\rho_{ij}| =\sum_\alpha |\rho_{ij}^{\alpha \alpha}|,$ which
results in an improved linear correlation matrix, that is more similar
to the NMI, see Fig.\ \SMCCA{a}.  More rigorously, we employ canonical
correlation analysis (CCA) as a standard tool of multivariate
statistical analysis, \cite{Hardoon04} which was previously used to
calculate the linear correlation between Cartesian
coordinates.\cite{Briki94} Introducing linear transformations
$A,B \in \mathbb{R}^{3\times 3}$ of the original coordinates $X$ and
$Y$, CCA calculates the canonical variables $\widehat X = AX$ and
$\widehat Y = BY$, which maximize the canonical correlation
\begin{equation} \label{eq:CCS}
\rho_{\rm C}(X,Y) \equiv \rho(\widehat X,\widehat Y) \stackrel{!}{=} {\rm max}.
\end{equation}
This leads to an eigenvalue problem for the canonical variables
$\widehat X$ and $\widehat Y$, which yields the canonical correlation
from a sum of the eigenvalues, i.e.,
$\rho_{\rm C}^2 = \frac13 \sum_i \lambda_i$. As shown in the SI, the
eigenvalues can be calculated directly from the $3\times 3$ correlation
matrices, which renders CCA a rather efficient method.
Comparing the resulting CCA matrix to the NMI, Fig.\ \SMCCA{b} reveals
good overall agreement of the two correlation measures. As first
  discussed by Lange and Grubm\"uller,\cite{Lange06} this shows that
the main discrepancy between the Pearson correlation and the NMI is
due to the lack of an appropriate treatment of the relative
orientation of the components, rather than intrinsically nonlinear
correlation effects.

%
%%%%%%%%%%%%%%%%%%%%%%%%%%%%%%%%%%%%%%%%%%%%%%%%%%%%%%%%%%%%%%%%%%%%%
%

As contour plots of similar correlation matrices look very much the
same, we change to an alternative representation to compare various
versions of the NMI. That is, we sort the $\sim 13~000$ residue pairs
in the order of increasing values of the (un-normalized) mutual
information $I$, such that $I$ is a monotonically increasing
function. Using this ordering, Fig.\ \ref{fig:T4L}e compares the
following variants of the NMI:
\begin{itemize}
\item $I_\text{NM}(X,Y) = I(X,Y)/I_{\rm max}$,
\item $I_{\rm N}(X,Y) = I(X,Y)/\sqrt{H(X)H(Y)}$ ,
\item $I_\text{NJ}(X,Y) = I(X,Y)/H(X,Y)$,
\item $I_\textsc{GY}(X,Y) = \sqrt{1-\exp (-I(X,Y)/3)}$,
\item and $\rho_\text{C}(X,Y)$ defined in
  Eq.\ \eqref{eq:CCS}.
\end{itemize}
The first three measures look quite similar, that is, they start at
$\sim 0.06$, exhibit a slow, approximately linear increase until
$\sim 12~000$, before they show a rapid rise to its final value of
1. That is, the vast majority (more than $\sim 90\%$) of residue pairs
show a correlation of $I/I_{\rm max}$ between $0.1$ and $0.3$. This is
in clear difference to $I_\textsc{GY}$ that is significantly shifted
to higher values such that the majority ($\sim 90\%$) of residue pairs
show a correlation between 0.5 and 0.8. This means, for example, that
two weakly correlated residues with $I_{\rm N} = 0.2$ yield
$I_\textsc{gy} \approx 0.7$, which seems to indicate considerable
correlation. Hence, the interpretation of $I_\textsc{gy}$ might be
misleading, although the limit for normally distributed variables is
given by the Pearson coefficient and therefore well defined. The
canonical correlation $\rho_\text{C}$, shows also rather high values
and significant vertical spread of the data. Presumably, the latter
finding is caused by nonlinear effects that are not described by the 
linear measure $\rho_\text{C}$.

We now focus on the first three measures. First we note that our
standard definition $I_{\rm N}$ follows the maximum-normalized
$I_\text{NM}$ very closely and shows a rather small average spread of
approximately $0.04$.  $I/H(X,Y)$ behaves quite similar, but is shifted to
smaller values due to Eq.\ \eqref{eq:upperlimits}. The overall
agreement of these two measures with $I/I_{\rm max}$ reveals that the
respective normalization factors $1/\sqrt{H(X)H(Y)}$ and $1/H(X,Y)$
are roughly constant and depend only little on the considered residue
pair. As shown in Eqs. \eqref{eq:Hcx_estimator} and
\eqref{eq:Hcxy_estimator}, this dependence enters solely via the
function $\epsilon$, representing the distance between a considered
data point and its $k$th neighbor. For a densely packed protein such
as T4L, $\epsilon$ appears to be quite similar for all residue pairs,
and causes only a small vertical spread of $I_{\rm N}$ and $I/H(X,Y)$.

The excellent agreement of $I_{\rm N}$ with $I/I_{\rm max}$ seems to
validate $I_{\rm N}$ as standard definition of the NMI. We note,
however, that the applicability of the maximum-normalized
$I/I_{\rm max}$ requires the existence of at least one perfectly
correlated data point (i.e., with $I_{\rm N} \approx 1$). In the case of
proteins, this is typically obtained for directly neighboring residues. In
contrast, when we use a data set with $I_{\rm N} \ll 1$ this agreement
breaks down. For example, by excluding next-neighbor correlations of
T4L, % such that $I_{\rm N} \le 3.5 $, 
the resulting maximum-normalized quantity $I_\text{max}^\text{NaN}$ 
shifts towards larger values (Fig.\ \ref{fig:T4L}e), and does no
longer agree with $I_{\rm N}$.

%\newpage
%
%%%%%%%%%%%%%%%%%%%%%%%%%%%%%%%%%%%%%%%%%%%%%%%%%%%%%%%%%%%%%%%%%%%%%
%
\subsection{Interresidue NMI vs contact correlation}
\vspace{-4mm}

We finally discuss the ability of the $C_\alpha$-based NMI
to explain the mechanism of the open-close
transition in T4L. Let us first recap previous results,\cite{Post22a}
which were based on the analysis of interresidue contacts. As
explained above, the open-close transition of the mouth region
(including the $\beta$-sheets of the N-domain and the helices
$\alpha_5$ and $\alpha_9$ of the C-domain) is triggered by the locking
motion of Phe4 in the hinge region (including the helices $\alpha_1$
and $\alpha_3$). The cooperative process is mediated via a network of
$\sim30$ highly correlated contacts,\cite{note3nmi} which range from the
hinge region to the mouth, see Fig.~\ref{fig:T4L2}a. Indicated by red
dots, Fig.~\ref{fig:T4L2}c shows that these contacts in particular
connect helices $\alpha_1$ and $\alpha_3$ in the hinge region, as well
as $\beta$-sheets $\beta_1$--$\beta_3$ and helices $\alpha_5$ and
$\alpha_9$ in the mouth region.  For example, during the open-closed
transition, in the hinge region the salt bridge between Glu5 and Lys60
opens, while in the mouth region the salt bridges Asp20-Arg145 and
Glu22-Arg137 close.

We note that correlated contacts identified by MoSAIC \cite{Diez22}
typically show a high difference of their population probabilities
$\Delta p = p_{\rm open} - p_{\rm closed}$ to exist in the open and the closed
state, see Tab.\ \SMTabMosaic{}. That is, 20 contacts exhibit
$\Delta p \gtrsim 0.7$, meaning that the contact approximately is open
in one state and closed in the other state. The remaining 12 MoSAIC
contacts showed $\Delta p \lesssim 0.2$, corresponding to a small (but
correlated) change of the contact distance. We also found $\sim 10$
contacts with moderate population changes (between 0.3 and 0.7) and
correlations (between 0.2 and 0.5).
While a large population change clearly indicate the participation of
a contact in an allosteric transition, \cite{Yao19b,Yao22} highly correlated
contacts as identified by Mosaic moreover indicate a cooperative process.

%\vspace*{-20mm}
\begin{figure}[ht!]
    \centering
    \includegraphics[width=0.9\linewidth]{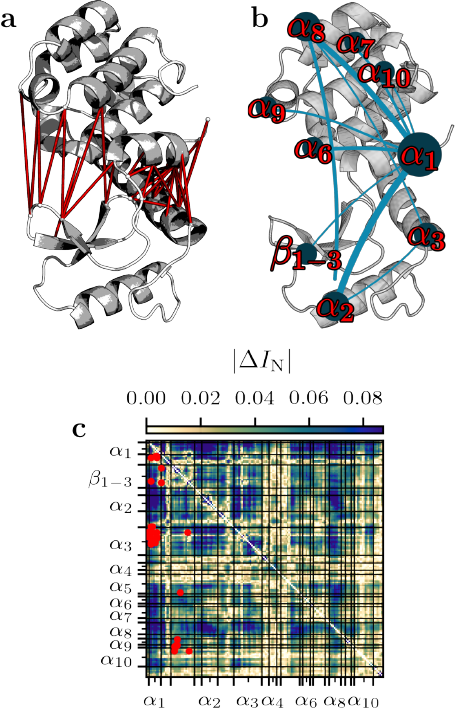}
    \caption{Illustration of the main structural correlations of T4L
      during the open-closed transition.  (a) Network of correlated
      inter-residue contacts identified by MoSAIC. \cite{Diez22} (b)
      Network representation of the most significant changes of the
      NMI, $\Delta I =I_\text{N}^\text{closed}-I_\text{N}^\text{open}$. The
      edge width reflects the average $|\Delta I_\text{N}|$ between
      two secondary structures, showing only edges with
      $|\Delta I_N| \geq 0.06$.  (c) Matrix representation of
      $|\Delta I|$, which also shows the MoSAIC contacts (red dots). }
    \label{fig:T4L2}
\end{figure}

To relate the above contact-based switching mechanism to the information
that can be learned from the $C_\alpha$-based NMI of T4L, it is
helpful to describe the open and the closed state separately and
consider the difference of the NMI in both conformations
\begin{equation} \label{eq:DI}
    \Delta I = I^\text{closed}_\text{N} - I^\text{open}_\text{N} ,
\end{equation}
the modulus of which is shown in Fig.~\ref{fig:T4L2}c.  Interestingly,
$\Delta I$ is found to overall decrease in the closed state (i.e.,
$\Delta I < 0$), except for the $\alpha_4$-helix separating the two
domains of T4L. As two rigidly connected residues are highly
correlated, while flexibly connected residues are less so, we can
typically associate high correlation with a relatively rigid
connection (since a rigid connection allows for an easy prediction of
the position of one part by knowing the position of the other).  That
is, $\Delta I_{ij} < 0$ means that the residues $i$ and $j$ are more
rigidly coupled in the open state. This is in line with the
root-mean-square-fluctuations of T4L (Fig.~\SMRMSF), which are overall
enhanced in the open state, except for helices $\alpha_4$ and
$\alpha_{10}$ whose fluctuations are similar. 
To facilitate the interpretation of the NMI difference matrix
$\Delta I_{ij}$, Fig.~\ref{fig:T4L2}b reveals a network representation
of the most important interactions (shown as blue patches in
Fig.~\ref{fig:T4L2}c). Restricting ourselves to
$|\Delta I_{ij}| \ge 0.06$, we find that (almost) all high NMI differences
involve the $\alpha_1$-helix.
%, except for the
%high differences of the outmost groups $\alpha_2$ and $\alpha_{8}$. 
In this sense, the blue lines that connect $\alpha_1$ to the N-domain
and the C-domain can be considered as allosteric pathways.

When we compare the coupling networks in Figs.\ \ref{fig:T4L2}a and
\ref{fig:T4L2}b, we see that both the contact-based representation and
the NMI difference matrix highlight the $\alpha_1$-helix (including
key residue Phe4) as the main driver of the open-closed process. While
the contact changes identified by MoSAIC account for the origin of the
conformation transition (i.e., the cooperative breaking and making of
contacts during the transition), the $C_\alpha$-based NMI differences
rather report on the change in rigidity of the system, that is, on the
impact of the transition.
\vspace{-4mm}

%\newpage
%
%%%%%%%%%%%%%%%%%%%%%%%%%%%%%%%%%%%%%%%%%%%%%%%%%%%%%%%%
%
\section{Concluding Remarks}
\vspace*{-4mm} 

We have generalized the widely used KSG-estimator\cite{Kraskov04} of
the mutual information (MI) to facilitate the calculation of the
normalized MI (NMI). The normalization is achieved by introducing the
relative entropy, which is invariant under variable transformations.
\cite{Jaynes68} Our new estimator of the NMI correctly reproduces results of a
histogram-based calculation for one-dimensional problems, but is beyond
that capable of treating multi-dimensional systems due to the use
of $k$-nn statistics.
The algorithm is numerically
efficient and does not require more effort than the calculation of
the (un-normalized) MI.

Considering the open-closed transition of T4L as a well-studied
example of allosteric coupling, we have calculated the NMI of the
C$_\alpha$-atom coordinates. Comparing various correlation measures,
we demonstrated that the $k$-nn-based estimation $I_\text{N}$
represents a correct upper bound of the NMI (assuming normalization
factors that contain both variables) and agrees well with the
maximum-normalized MI $I/I_{\rm max}$ (if the data contain at least
one perfectly correlated point). Conversely, we showed that the
popular rescaled measure $I_\textsc{GY} = \sqrt{1-\exp (-I/3)}$
typically overestimates the correlation significantly. We also
showed that the linear Pearson correlation coefficient
$\rho_{ij}$ fails completely to reproduce the correct NMI of Cartesian
coordinates, while the linear canonical correlation analysis
\cite{Hardoon04,Briki94} was found to yield qualitatively correct
results at low computational cost.
Finally we discussed the merits of the $C_\alpha$-based NMI to explain
the structural mechanism of the open-closed transition of T4L.
\vspace*{-4mm}

%
%%%%%%%%%%%%%%%%%%%%%%%%%%%%%%%%%%%%%%%%%%%%%%%%%%%%%%%%%
%

\subsection*{Supplementary material}
\vspace*{-4mm} 
Supplementary methods include the description of the radius-based
algorithm for the volume estimator (Fig.~\SMmeanradius),
the runtime comparison (Fig.~\SMruntime) and the canonical
correlation analysis (Fig.\ \SMCCA). Supplementary results for T4L
include rotational fit analysis (Fig.~{\SMRotfit}), spatial
components of the Person correlation (Fig.~\SMxyz), the RMSF
(Fig.~\SMRMSF), and Tab.\ \SMTabMosaic{} listing important interresidue
contacts.
\vspace*{-6mm} 

\subsection*{Acknowledgments}
\vspace*{-4mm}
The authors thank Steffen Wolf and Nele Dethloff for helpful comments 
and discussions. This work
has been supported by the Deutsche Forschungsgemeinschaft (DFG) within
the framework of the Research Unit FOR 5099 ''Reducing complexity of
nonequilibrium'' (project No.~431945604), the High Performance and
Cloud Computing Group at the Zentrum f\"ur Datenverarbeitung of the
University of T\"ubingen, the state of Baden-W\"urttemberg through
bwHPC and the DFG through grant no INST 37/935-1 FUGG (RV bw16I016),
and the Black Forest Grid Initiative.  
\vspace*{-4mm}

\subsection*{Data Availability Statement}
\vspace*{-4mm} 
The estimator is freely available at \url{https://github.com/moldyn/normi},
or via \texttt{pip/conda install normi}, and adapts the scikit-learn
syntax.\cite{Pedregosa11}

%\newpage
\bibliography{\dir/stock,\dir/md,new}

\begin{thebibliography}{10}
\bibitem{note0nmi}
{In taking the limit $N\to\infty$, Jaynes encountered an additional diverging
  term $\ln N$, which is usually omitted \cite{Jaynes68}}.
\bibitem{note4nmi}
{To minimize bias due to different radii in the joint and marginal spaces, they
  estimate marginal entropies not with a fixed $k$ but with a fixed
  $\varepsilon$ taken from the joint entropy. So if the $\varepsilon_i$ ball at
  $x_i$ contains $n_{x_i}$ points, the distance corresponds to the $n_{x_i} +
  1$ nearest-neighbor distance}.
\bibitem{note5nmi}
{As in Jaynes' derivation of the relative entropy,\cite{Jaynes68} we neglect
  the term $\ln N$.}
\bibitem{note3nmi}
{The number of $\sim 30$ correlated contacts obtained from MoSAIC for a
  resolution parameter $\gamma = 0.5$ agrees with the results obtained in
  Ref.~\citenum{Diez22}, but differs to the results of Ref.~\citenum{Post22a}
  that mistakenly included hydrogen atoms in the contact definition, which
  resulted in $\sim 85$ correlated contacts}.
\end{thebibliography}

\end{document}

\title{Supplementary Material for:\\
Accurate estimation of the normalized mutual information of multidimensional data}

\author{Daniel Nagel}
\affiliation{Biomolecular Dynamics, Institute of Physics, University of Freiburg, 79104 Freiburg, Germany}
\email{stock@physik.uni-freiburg.de}
\author{Georg Diez}
\affiliation{Biomolecular Dynamics, Institute of Physics, University of Freiburg, 79104 Freiburg, Germany}
\email{stock@physik.uni-freiburg.de}
\author{Gerhard Stock}
\affiliation{Biomolecular Dynamics, Institute of Physics, University of Freiburg, 79104 Freiburg, Germany}
\email{stock@physik.uni-freiburg.de}
\date{\today}

\maketitle

\baselineskip5.4mm

\section{Alternative Invariant Measures}
\begin{figure}[ht!]
    \includegraphics{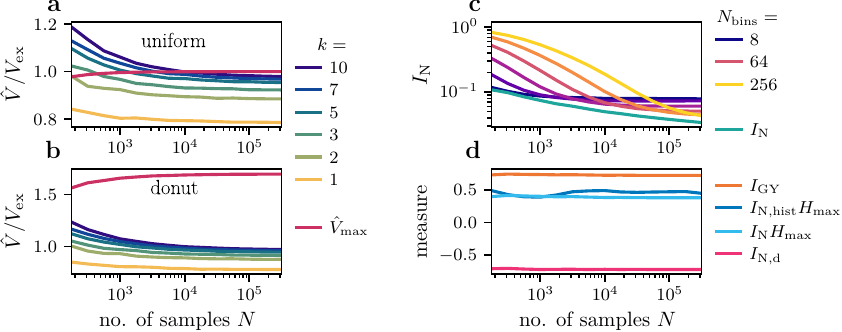}
    \caption{
    \setlength{\baselineskip}{4mm}
    Estimating the 2d volume via Eq.~\eqref{SI:eq:volume_r}
    (a+b) Estimation of the normalized volumes of various 2D
    distributions, using the $k$-nn estimator
    [Eq.~\eqref{SI:eq:volume_r}, $k\in[1,10]$] and the boundary estimate
    [Eq.~\EQjaynesassumption, red lines]. Shown as a
    function of sample size $N$, the results are shown for (a) a
    uniform distribution ($x, y\in [0, 1]$) and (b) uniformly distributed
    data in a donut-shaped area $r=\sqrt{x^2+y^2}\in[0.5, 1]$.
    (c+d) Accuracy of the new
    $k$-nn estimator of the NMI [Eq.~\EQminormalization],
    obtained for a 2D toy model [Eq.~\EQtwodmodel].
    (c) Comparison of $k$-nn estimation of the NMI,
    shown as green line, with the discrete NMI estimation (2D binning)
    of the data are shown (colored lines) for various numbers of bins.
    (d)
    Convergence comparison of the MI estimator $I$
    [Eq.~\EQksgdifferential], the $I_\text{GY}$
    [Eq.~\EQigy], and the NMI estimators based on the
    relative entropy $I_\text{N}$ [Eq.~\EQhcxestimator], 2D binning
    $I_\text{N, hist}$, and the differential entropy $I_\text{N,D}$
    [Eq.~\EQksgentropyestimator]. $H_\text{max}$ corresponds
    to the highest possible entropy, that is, the entropy of a uniform
    distribution, for a given number of samples $N$.}
  \label{SI:fig:volume}
\end{figure}
As mentioned in the main paper, one could also estimate the volume of
by relying on the mean $k$-nn radius, namely
\begin{align}
    \hat{V}_r(X,Y) &= \frac{N}{k} c_{d_X} c_{d_Y}\langle 2 \varepsilon \rangle^{d_X + d_Y}\;, \label{SI:eq:volume_r}
\end{align}
where $N$ is the number of data points, $\varepsilon$ the $k$-th
nearest neighbor distance, $c_d$ the volume element, and $d$ the
dimensionality. In Fig.~\ref{SI:fig:volume}a,b the estimator is
applied to the two examples discussed in the main paper, see
Fig.~\Figone. While all estimators converge rapidly,
we find that for small $k$ there is a systematic bias to underestimate
the volume.

Following the arguments of the main paper to ensure correct
factorization, we define the partial volumes by
\begin{align}
        \hat{V}_r(X,Y) &= \hat{V}_r(X)\hat{V}_r(Y)\\
        \Rightarrow\; \hat{V}_r(X) &= \sqrt{\frac{N}{k}} c_{d_X} \langle 2 \varepsilon\rangle^{d_X}\;.
\end{align}
Therewith, we get directly the invariant measures
\begin{align}
    \ln m_r(X) &= - \ln c_{d_X} - d_X\ln \langle 2\varepsilon \rangle, \\
    \ln m_r(X, Y) &= - \ln (c_{d_X} c_{d_Y}) - (d_X + d_Y)\ln \langle 2\varepsilon \rangle\label{eq:m_r}\;.
\end{align}
To define the entropy estimators Eq.~\EQhcxestimator~and Eq.~\EQhxhyestimator
this results in a rescaled $k$-th nearest neighbor radius of
\begin{align}
    \tilde{\varepsilon}_r &= \frac{\varepsilon}{\langle \varepsilon\rangle} \;.
\end{align}
In Fig.~\ref{SI:fig:volume}c,d we compare the effect of the volume estimation on the
NMI estimator $I_N$. It is clearly visible that for a distribution without fixed
supports the border effects are negligible and results are in good agreement
with the other volume definition, see Fig.~\Figtwo{} in the main paper.

\section{Runtime Comparison}
\begin{figure}[ht!]
    \includegraphics{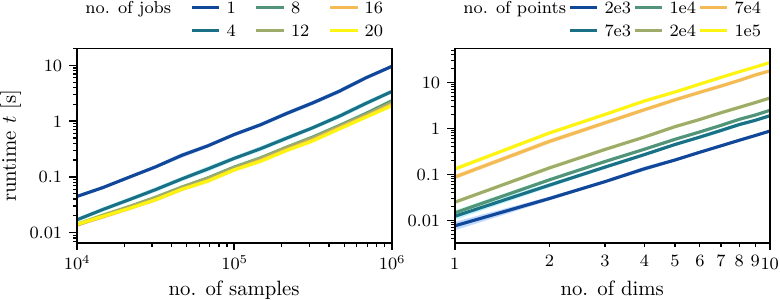}
    \caption{
    \setlength{\baselineskip}{4mm}
    Runtime comparison of the here presented NMI estimator on
    synthetic data $\boldsymbol{x}, \boldsymbol{y}$ using an
    Intel\textsuperscript{\textregistered{}} Core\texttrademark{}
    i9-10900 CPU.
    (left) Runtime depending on the no. of samples and jobs for 1D
    dimensional data, and (right) runtime depending on the no. of
    dimensions and samples.
    In the case of protein dynamics, this could result in a runtime of
    $\SI{\approx24}{\hour}$ ($\SI{\approx1}{\hour}$) for $1100$ ($230$)
    backbone dihedral angles ($d=1$) using $N=10^5$ frames. In case of
    Cartesian coordinates ($d=3$), this could result in a runtime of
    $\SI{\approx24}{\hour}$ ($\SI{\approx1}{\hour}$) for $900$ ($190$)
    residues using $N=10^4$ frames.
    Data are drawn randomly according to $x_i = \mathcal{N}(\mu=\pm1,
    \sigma^2=1)$, where $\mathcal{N}$ refers to a normal distribution,
    and the sign alternates for successive samples, and $y_i = x_i +
    0.2\:\mathcal{N}(\mu=0, \sigma^2=1)$.
    For this benchmark we used $k=3$.}
    \label{SI:fig:runtime}
\end{figure}

\clearpage
\section{Rotational fit}
\begin{figure}[h!]
    \centering
    \includegraphics[width=0.45\linewidth]{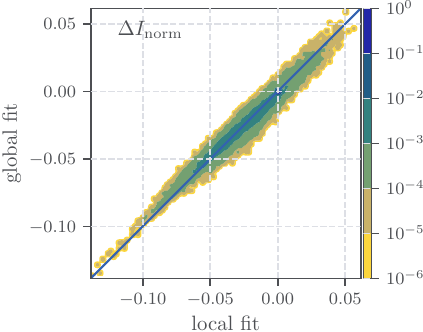}
    \caption{
        \setlength{\baselineskip}{4mm}
        Comparison of the relative differences between the local and the
        global fitting procedure for T4L. Shown is the joint probability
        distribution of the mutual information difference
        $\Delta I_\text{N} = I_\text{N}^\text{closed} - I_\text{N}^\text{open}$
        computed using a local and global fit.
        In the local fitting procedure, the collective rotation and
        translation effects were eliminated by executing a root mean
        square deviation (RMSD) fit to the structure featuring the
        minimal average RMSD concerning all other frames within
        each distinct conformation (open and closed).
        Conversely, in the global fitting approach, the frame
        characterized by the minimum average RMSD relative to the
        complete trajectory was employed.
    }
    \label{SI:fig:fitting}
\end{figure}

\FloatBarrier
\section{Linear Pearson correlation of T4L}
\begin{figure}[h!]
    \includegraphics[width=1.0\linewidth]{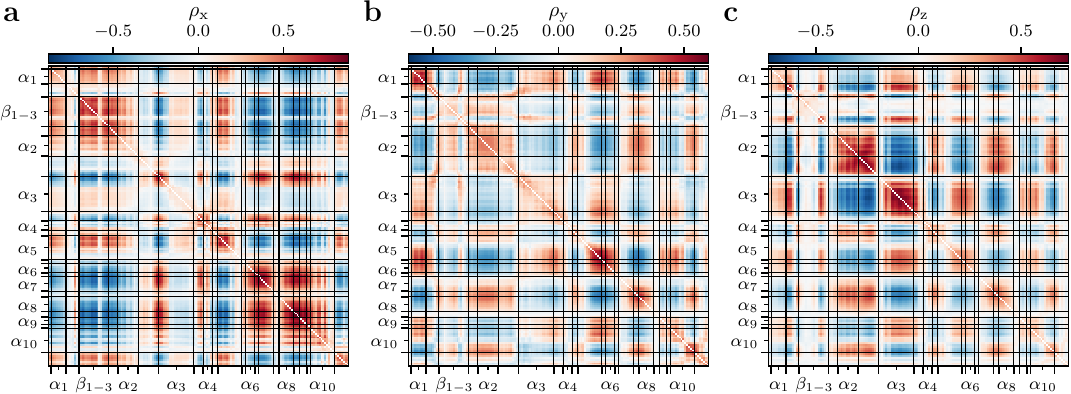}
    \caption{
    \setlength{\baselineskip}{4mm}
    The contributions of the individual directions $x, y$
      and $z$ to the linear correlation between the Cartesian
      coordinates of the C$_\alpha$-atoms of T4 lysozyme.  Focusing on
      the relationship between, e.g., the $\alpha_6$ and the
      $\alpha_8$ helix, there is a large difference between the
      Cartesian correlation and normalized mutual information (compare
      Fig.~\Figthree{}a and c).  This can be explained by the fact
      even though this region is correlated in all directions, the
      sign differs. It is positive for the $x$-direction and negative
      for the $y$- and $z$-directions which results in a cancelation of
      the correlation.  }
    \label{SI:fig:CartCorr_xyz_seperately}
\end{figure}

\FloatBarrier
\section{Canonical Cartesian Correlation}
\begin{figure}[ht!]
    \centering
    \includegraphics[width=0.8\linewidth]{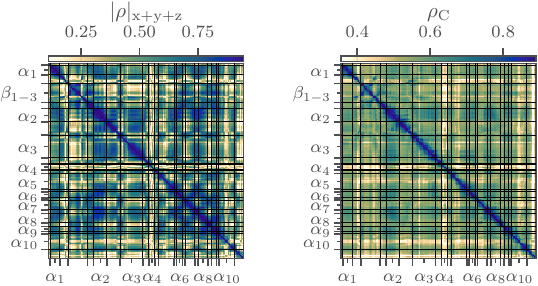}
    \caption{
    \setlength{\baselineskip}{4mm}
    Comparing linear Cartesian correlation measures for T4L.
    On the left-hand side, the absolute correlation contributions of each
    direction (see Fig.~\ref{SI:fig:CartCorr_xyz_seperately}) are summed up,
    while on the right-hand side the canonical correlation is shown.}
    \label{SI:fig:cancorr_corr}
\end{figure}

The Pearson correlation coefficient has the major drawback of not respecting
correlations between different directions in the Cartesian space $i$ and $j$
when $i\neq j$, $i,j \in \{x,y,z\}$.
Therefore, even strongly linear couplings
between two sets of multidimensional variables can not be resolved
(see Fig.~\ref{SI:fig:CartCorr_xyz_seperately}).
As a remedy, canonical correlation analysis tries to identify linear
transformations that transform both variables in such a way that the
resulting \textit{canonical variables} feature maximum correlation.
It can be shown,\cite{Briki94} that the canonical correlation coefficient can be
calculated as
\begin{align}
    &\rho_\text{C} = \sqrt{\frac{1}{3} \text{tr}({R})},\\
    \intertext{where $R$ is given by}
    &R = R_{11}^{-1} R_{12} R_{22}^{-1} R_{21},
    \intertext{and $R_{ij}$ can be calculated as}
    &R_{nm} = \begin{pmatrix}
        \langle \delta x_n \delta x_m \rangle &  \langle \delta x_n \delta y_m \rangle & \langle \delta x_n \delta z_m \rangle \\
        \langle \delta y_n \delta x_m \rangle &  \langle \delta y_n \delta y_m \rangle & \langle \delta y_n \delta z_m \rangle \\
        \langle \delta z_n \delta x_m \rangle &  \langle \delta z_n \delta y_m \rangle & \langle \delta z_n \delta z_m \rangle \\
    \end{pmatrix}.
\end{align}
Here, $n, m \in \{1, 2\}$ denote the sets of variables $1$ and $2$, that is, the
Cartesian coordinates of two C$_\alpha$-atoms and
$\delta \alpha$ are standard normalized variables $\delta \alpha = \alpha - \langle \alpha \rangle / \sqrt{\langle ( \alpha - \langle \alpha \rangle)^2\rangle}$, where
$\alpha$ can denote $x, y$ and $z$.
The resulting canonical correlation matrix for T4L is shown in
Fig.~\ref{SI:fig:cancorr_corr} on the right-hand side.
For comparison, the sum of the absolute contributions in
$x$, $y$ and $z$ (shown in Fig.~\ref{SI:fig:CartCorr_xyz_seperately})
directions is shown in Fig.~\ref{SI:fig:cancorr_corr}
on the left-hand side.

\FloatBarrier
\section{Flexibility: Root-Mean-Square-Fluctuations}
\begin{figure}[ht!]
    \includegraphics[width=1\linewidth]{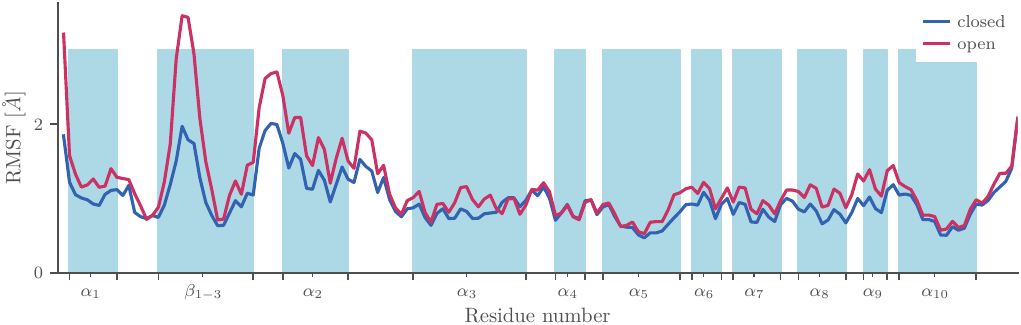}
    \caption{
        \setlength{\baselineskip}{4mm}
        The root-mean-square-fluctuation (RMSF)
        $\sqrt{\langle \left( \mathbf{x}_i - \langle \mathbf{x}_i
        \rangle \right) ^2 \rangle}$ as a function for every C$_\alpha$-atom
        in the system. The RMSF was calculated for both the closed
        and open conformation. While the lower part of T4L
        ($\alpha_1 - \alpha_2$) shows a larger difference in RMSF
        in both conformations, the upper part seems more rigid.
        This is especially the case for the $\alpha_4$ and $\alpha_{10}$
        helices.
    }
    \label{SI:fig:RMSF}
\end{figure}

\bibliography{Bib/stock.bib,Bib/new.bib}

\begin{table}[]
\centering
\renewcommand{\arraystretch}{0.6}
\begin{tabular}{l|llll}
Contacts & $p_\text{open}$ & $p_\text{closed}$ & $\Delta p$ & $\langle |\rho |\rangle_{\text{C}_1}$ \\
\hline
$r_{4,60}$ & 0.94  & 0.00 & 0.94 & 0.82 \\
$r_{4,63}$ & 0.92  & 0.00 & 0.92 & 0.81 \\
$r_{4,13}$ & 0.91  & 0.00 & 0.91 & 0.81 \\
$r_{4,29}$ & 0.96  & 0.00 & 0.96 & 0.78 \\
$r_{22,137}$ & 0.00  & 0.89 & 0.88 & 0.77 \\
$r_{4,64}$ & 1.00  & 0.06 & 0.93 & 0.76 \\
$r_{8,67}$ & 0.00  & 0.96 & 0.96 & 0.76 \\
$r_{22,141}$ & 0.00  & 0.83 & 0.83 & 0.76 \\
$r_{21,141}$ & 0.00  & 0.80 & 0.80 & 0.76 \\
$r_{2,64}$ & 0.52  & 0.00 & 0.52 & 0.76 \\
$r_{7,71}$ & 0.01  & 0.97 & 0.96 & 0.74 \\
$r_{4,71}$ & 0.00  & 0.92 & 0.92 & 0.72 \\
$r_{21,142}$ & 0.00  & 0.59 & 0.59 & 0.71 \\
$r_{7,12}$ & 1.00  & 0.17 & 0.82 & 0.70 \\
$r_{8,64}$ & 0.00  & 0.84 & 0.84 & 0.69 \\
$r_{4,68}$ & 0.00  & 0.84 & 0.84 & 0.66 \\
$r_{8,13}$ & 0.97  & 0.21 & 0.76 & 0.65 \\
$r_{3,67}$ & 0.98  & 0.13 & 0.85 & 0.64 \\
$r_{8,12}$ & 1.00  & 0.26 & 0.73 & 0.62 \\
$r_{11,30}$ & 0.26  & 0.92 & 0.66 & 0.57 \\
\hline
$r_{4,72}$ & 0.00  & 0.12 & 0.12 & 0.77 \\
$r_{20,142}$ & 0.00  & 0.08 & 0.08 & 0.77 \\
$r_{8,68}$ & 0.00  & 0.05 & 0.05 & 0.76 \\
$r_{1,64}$ & 0.05  & 0.00 & 0.05 & 0.72 \\
$r_{30,145}$ & 0.00  & 0.05 & 0.05 & 0.72 \\
$r_{5,60}$ & 0.15  & 0.00 & 0.15 & 0.71 \\
$r_{20,145}$ & 0.00  & 0.09 & 0.09 & 0.70 \\
$r_{5,64}$ & 0.04  & 0.00 & 0.04 & 0.64 \\
$r_{24,105}$ & 0.00  & 0.08 & 0.08 & 0.63 \\
$r_{29,64}$ & 0.00  & 0.12 & 0.12 & 0.60 \\
$r_{11,20}$ & 0.03  & 0.26 & 0.24 & 0.58 \\
$r_{2,67}$ & 0.03  & 0.00 & 0.03 & 0.58 \\
\hline
$r_{75,88}$ & 0.85  & 0.31 & 0.54 & 0.48 \\
$r_{11,18}$ & 0.17  & 0.85 & 0.68 & 0.46 \\
$r_{3,75}$ & 0.00  & 0.32 & 0.32 & 0.45 \\
$r_{10,104}$ & 0.00  & 0.35 & 0.35 & 0.42 \\
$r_{7,100}$ & 0.12  & 0.60 & 0.47 & 0.38 \\
$r_{29,104}$ & 0.58  & 0.98 & 0.40 & 0.38 \\
$r_{84,103}$ & 0.26  & 0.81 & 0.56 & 0.37 \\
$r_{31,69}$ & 0.44  & 0.07 & 0.37 & 0.34 \\
$r_{104,145}$ & 0.18  & 0.52 & 0.34 & 0.29 \\
$r_{14,20}$ & 0.36  & 0.02 & 0.34 & 0.21 \\
$r_{81,108}$ & 0.53  & 0.84 & 0.30 & 0.13 \\
\end{tabular}
\caption{
    \label{tab:contacts}
    \setlength{\baselineskip}{4mm}
    List of 43 inter-residue contacts that mediate the open$\leftrightarrow$closed
    transition of T4L:
    (Top:)
    $20$ highly correlated contacts that are most
    important for the open-closed transition as they are highly correlated
    $\langle |\rho| \rangle_{\text{C}_1}$ in the MoSAIC\cite{Diez22} analysis of cluster 1
    and feature a high change in contact probability $\Delta p = | p_\text{open} - p_\text{closed}|$.
    (Middle):
    12 contacts, that are also highly correlated,
    but exhibit contact probability changes $\Delta p_\text{C} \leq 0.3$.
    (Bottom): 11 contacts
    that feature a high contact probability change $\Delta p_\text{C} \geq 0.3$,
    but are not significantly correlated to the cordinates describing the
    open-closed transition $\langle |\rho| \rangle_{\text{C}_1} \leq 0.5$.
    }
\end{table}